# A non-equilibrium strategy for the general synthesis of single-atom catalysts

Yue Li,[1,2,§] Yang Xu,[1,§] Yunbiao Zhao,[1] Mingwei Cui,[2] Xiner Chen,[1] Liu Qian,[2] Jin Zhang,[2] Xueting Feng,[3]* Ziqiang Zhao[1,2]*

[1] State Key Laboratory of Nuclear Physics and Technology, School of Physics, Peking University, Beijing, 100871, China.

[2] School of Materials Science and Engineering, Peking University, Beijing, 100871, China.

[3] Department of Chemistry, Capital Normal University, Beijing, 100048, China.

[§] These authors contributed equally to this work: Yue Li and Yang Xu

* E-mail: zqzhao@pku.edu.cn; fengxt@cnu.edu.cn

**Abstract**

Single-atom catalysts (SACs) maximize atom efficiency and exhibit unique electronic structures, yet realizing precise and scalable atomic dispersion remains a key challenge. Here, we report a non-equilibrium strategy for the scalable synthesis of SACs via ion implantation, enabling precise stabilization of metal atoms on diverse supports. Using an industrial-grade ion source, wafer-scale ion implantation with milliampere-level beam currents enables high-throughput fabrication of SACs, while the synergistic energy-mass effects stabilize isolated metal atoms *in situ*. A library of 36 SACs was constructed, and the resulting Pt/MoS$_2$ exhibits outstanding hydrogen evolution performance with an overpotential of only 26 mV at 10 mA cm$^{-2}$ and exceptional long-term stability, surpassing commercial Pt/C. This work demonstrates ion implantation as a versatile platform bridging fundamental SACs design and scalable manufacturing, providing new opportunities for high-performance catalysts in energy conversion applications.

**Introduction**

Catalysis has been a cornerstone of chemical research and industrial processes due to its ability to facilitate energy-efficient and cost-effective chemical transformations. Among various advancements in catalysis, the development of single-atom catalysts (SACs) in the past decade has emerged as a frontier for catalysis, demonstrating superior activity compared to their bulk counterparts[1,2]. SACs are characterized by isolated metal atoms dispersed on suitable support materials, providing unique electronic structures and maximum surface exposure for catalytic processes, making them particularly advantageous for reactions such as hydrogen evolution[3,4], CO oxidation[1], and methane activation[5,6]. The synthesis protocols are pivotal in determining the structural, electronic, and consequently catalytic properties of SACs. Conventional high-temperature strategies (e.g., pyrolysis[7-11], chemical vapor deposition[12], atomic layer deposition[13]) often drive the system toward thermodynamic equilibrium, where metal atoms readily diffuse and, more frequently than achieving uniform atomization, aggregate into clusters to minimize the system's free energy[14-17]. In contrast, room-temperature strategies such as photochemical reduction[18,19] and electrochemical deposition[20,21] can circumvent thermal diffusion, yet the weak metal-support interactions formed under these mild conditions often yield poorly anchored species with limited stability under practical operation[22]. To overcome these challenges, researchers have focused on designing supports with sufficient defect sites[23] or tailored active sites[24,25] to stabilize the metal atoms and maintain their dispersion, even under elevated temperatures (Fig. 1a). Despite these efforts, sintering and atomization are antagonistic effects inherent to thermodynamic processes, making precise control challenging in practical applications.

A more fundamental strategy is to shift away from thermodynamic equilibrium and employ methods that emphasize kinetic control. Notably, ion implantation offers a non-equilibrium approach widely used in material modification and doping, with its most widespread application being in carrier doping for integrated circuits, where it is a mature industrial technology due to its precision and reliability[26,27]. By utilizing energetic ion beams to introduce atoms into substrates, it provides precise control over atomic distribution without necessitating high temperatures[28]. Due to these inherent advantages, ion implantation has been explored in catalyst development. For instance, several studies have highlighted its effectiveness in modifying nanostructured energy materials, achieving enhanced catalytic properties through defect and dopant introduction[29-31]. Controlled ion irradiation has also been shown to improve the catalytic activity of $MoS_2$ by generating sulfur vacancies, which optimize hydrogen adsorption sites[32,33]. While these studies provided valuable insights, the catalytic performance has not yet reached the levels seen with traditional chemical approaches. Moreover, their approaches often included post annealing, which reduced the benefits of ion implantation as a non-equilibrium process. The application of ion implantation in catalyst development still requires further exploration, both in terms of theoretical understanding and process engineering, to fully exploit its potential.

Herein, we exploit the non-equilibrium nature and scalability of ion implantation to develop a versatile strategy for SACs synthesis. Using an industrial-grade Metal Vapor Vacuum Arc (MEVVA) ion source, we achieved 8-inch wafer-scale beam spots with milliampere-level beam currents, enabling high-throughput fabrication of SACs. The ion implantation process provides two synergistic functions: energy effects that induce structural rearrangements[34] and generate new anchoring sites *in situ*, and mass effects that directly introduce metal atoms into the supports (Fig. 1b). By tuning the ion energy and dose, we effectively balanced these effects to stabilize dispersed metal atoms. On this basis, we established a library of SACs across four supports—molybdenum disulfide ($MoS_2$), graphene nanosheets (GN), graphdiyne (GDY) and carbon nanotubes (CNT)—and seven metals (Fe, Co, Ni, Cu, Pt, Ru, Mo). Their catalytic performance in hydrogen evolution was systematically evaluated, with Pt/$MoS_2$ delivering an overpotential of only 26 mV at 10 mA $cm^{-2}$ and superior stability to commercial Pt/C. These results highlight ion implantation as a universal and powerful platform for next-generation SACs design.

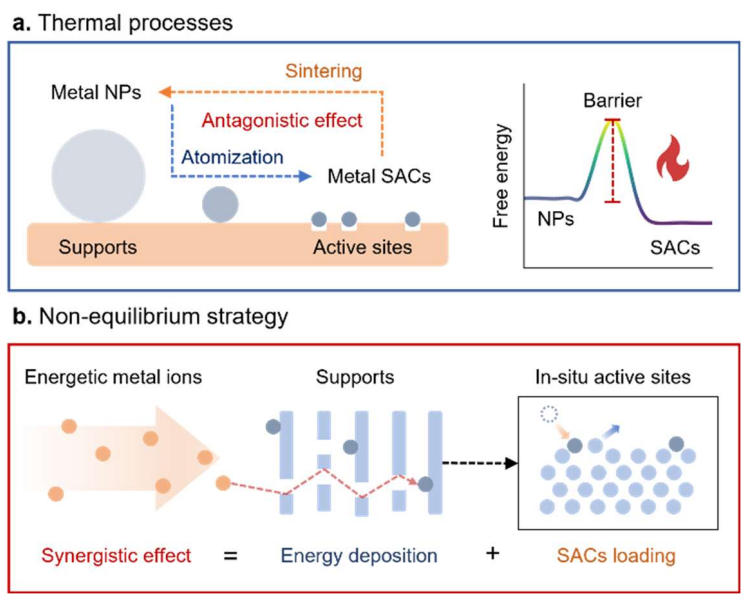

**Fig. 1. Comparison of mechanisms for synthesizing single-atom catalysts (SACs) using non-equilibrium strategy versus thermal processes. a,** Schematic illustration of the 'high-temperature paradox': while elevated temperatures promote metal anchoring, they simultaneously induce atomic aggregation, hindering the controlled synthesis of SACs. **b,** Mechanistic illustration of the ion implantation approach: the synergistic mass and energy effects of ion beams act on low-dimensional materials, enabling the formation of stably anchored single atoms.

## Results and discussion

### Synthesis and structure characterization of SACs

We synthesized a library of SACs using a room-temperature ion implantation method (Supplementary Fig. 1). Various combinations of metals (Fe, Co, Ni, Cu, Pt, Ru, Mo) and supports ($MoS_2$, GN, GDY, CNT) were obtained (Fig. 2a). To validate this strategy, Pt single atoms supported on $MoS_2$ ($Pt/MoS_2$), prepared under the optimized implantation conditions (10 keV, $5 \times 10^{16}$ ions cm$^{-2}$), were selected as a representative model system. Morphologically, the as-prepared $Pt/MoS_2$ shows same nanosheet structure of pristine $MoS_2$, with no observable Pt clusters or nanoparticles (Fig. 2b and Supplementary Fig. 2). Furthermore, no diffraction peaks attributable to crystalline Pt species were detected in the X-ray diffraction (XRD) patterns, with all observed peaks indexed to the $MoS_2$ support (Supplementary Fig. 3). Energy-dispersive X-ray spectroscopy (EDX) confirms the uniform dispersion of Pt, Mo, and S elements throughout the nanosheets (Fig. 2b). The Pt loading in the Pt/MoS$_2$, as determined by inductively coupled plasma optical emission spectrometer (ICP-OES), is 11.3 wt%. Importantly, the metal loading can be precisely regulated by adjusting the ion dose. Aberration-corrected high-angle annular dark-field scanning transmission electron microscopy (HAADF-STEM) image (Fig. 2c) shows the isolated Pt single atoms anchored on the MoS$_2$ lattice. Further inspection reveals that Pt atoms occupy two distinct sites: Pt substituting the Mo site (Pt$_{sub}$) and Pt adsorbed on the top site of S (Pt$_{ads}$-S)[35], confirming the single-atom nature of the implanted species.

The electronic properties and coordination environment of Pt were further investigated by X-ray absorption near-edge structure spectroscopy (XANES) and extended X-ray absorption fine-structure spectroscopy (EXAFS). As shown in Fig. 2d, the white line intensity of Pt L$_3$-edge XANES of Pt/MoS$_2$ lies between that of PtO$_2$ and Pt foil, suggesting that the Pt species in the Pt/MoS$_2$ are mainly in the oxidized state. The Fourier transformed $k^3$-weighted EXAFS spectrum of Pt/MoS$_2$ (Fig. 2e) shows a distinct peak at approximately 1.87 Å, assigning to the Pt-S contribution.[35] No obvious metallic Pt-Pt coordination peak is observed. X-ray photoelectron spectroscopy (XPS) was employed to investigate the valence states of Pt (Supplementary Fig. 4). The deconvoluted Pt 4f$_{7/2}$ peaks of Pt/MoS$_2$ appear at 72.5, 73.4, and 74.4 eV, corresponding to Pt$^{\delta+}$, Pt$^{2+}$, and Pt$^{4+}$, respectively[35], indicating the formation of Pt-S bonds. The wavelet transform (WT)-EXAFS analysis (Fig. 2f and Supplementary Fig. 5) also reflects the absence of Pt aggregation. The above results directly confirm the atomic-level dispersion of Pt in MoS$_2$.

Fourier transformed (FT)-EXAFS analysis and corresponding fitting results yielded a Pt-S bond length of 2.42 Å and a coordination number (CN) of 5 in the Pt/MoS$_2$ as shown in Supplementary Fig.6 and Supplementary Table 1. The intermediate CN falls between the typical values for substitutional and adsorbed Pt, suggesting the coexistence of both coordination environments (Pt$_{sub}$ and Pt$_{ads}$-S), which is consistent with the HAADF-STEM results. To verify that the above structural and electronic conclusions are not specific to Pt/MoS$_2$, other representative SACs were further characterized. Aberration-corrected HAADF-STEM and EDX mapping confirmed the uniform distribution of isolated metal sites across this broad library (Fig. 2g-j, Supplementary Figs. 7-10).

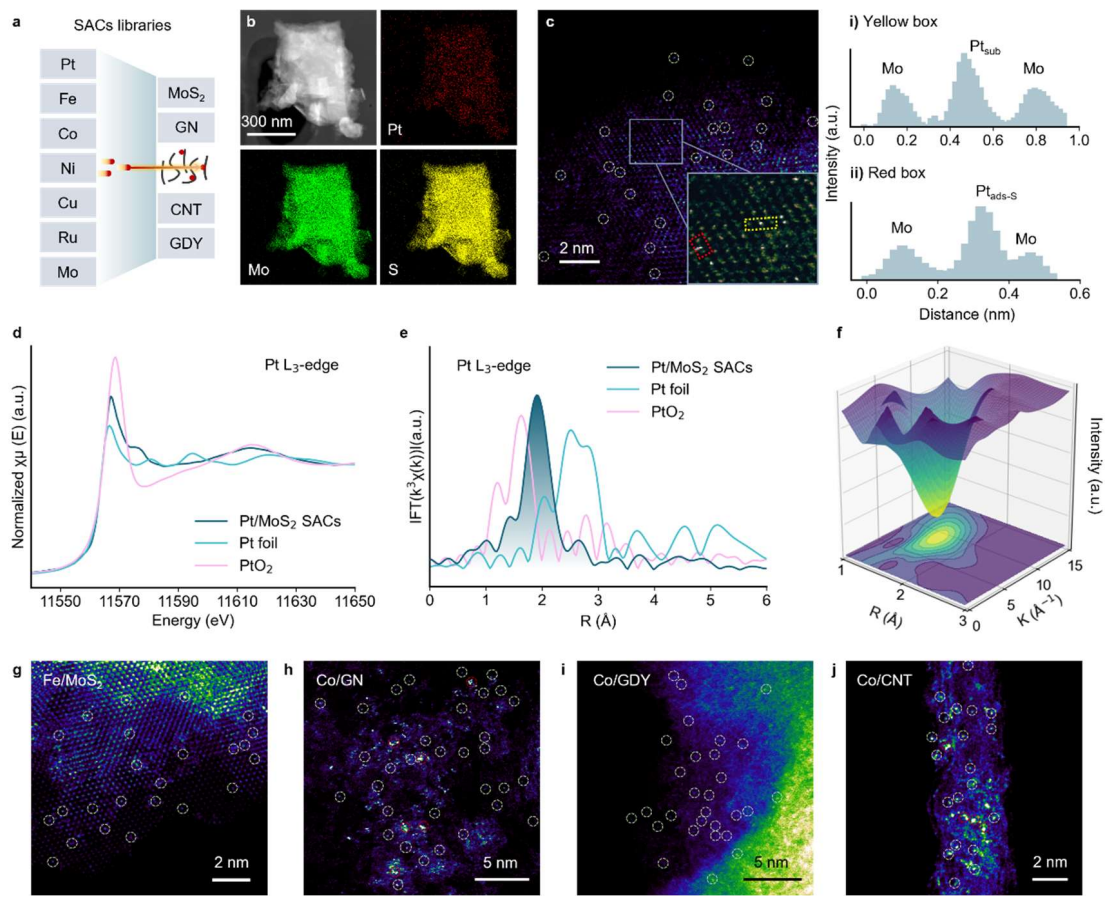

**Fig. 2. Visualization and spectroscopic characterization of SACs. a,** The metal elements and support used for preparing SACs. **b,** STEM image and corresponding EDX elemental mapping images of Pt/MoS$_2$. **c,** Aberration-corrected HAADF-STEM image of Pt/MoS$_2$. **d,** XANES spectra, **e-f,** FT-EXAFS spectra (e) and WT-EXAFS spectra (f) of Pt/MoS$_2$, referenced Pt foil and PtO$_2$. **g-j,** Aberration-corrected HAADF-STEM images of Fe/MoS$_2$ (g), Co/GN (h), Co/GDY (i), and Co/CNT (j).

## Synergistic mechanism of ion implantation

The successful synthesis of SACs (Fig. 2) validates the mass effects of the ion implantation strategy. Concurrently, these energetic ions inevitably deposit kinetic energy as they traverse and stop within the support. We therefore conducted a detailed investigation into the consequences of this energy effects, the non-equilibrium modification of the support structure. HAADF-STEM imaging (Fig. 3a) reveals that this ion bombardment induces atomic rearrangement, specifically the generation of monovacancies and divacancies. These *in situ* vacancies, with their highly unsaturated coordination, function as the high-activity "*in-situ* active sites" (Fig. 1b) that provide robust chemical anchoring for implanted metal atoms[36,37].

More strikingly, the ion beam's energy deposition was found to induce phase transitions in local regions of the support, converting the material from the 2H to the 1T' phase (Fig. 3b)[38-40]. This structural transformation is clearly resolved in the HAADF-STEM image by the formation of characteristic zigzag Mo-Mo chains and a distorted octahedral lattice, which is consistent with the atomic structure of 1T'-$MoS_2$. This metallic 1T' polymorph is recognized as an exceptional support for atomically dispersed Pt, offering higher conductivity and intrinsic hydrogen evolution reaction (HER) activity than the 2H phase [41-43].

To establish a quantitative correlation between these structural modifications and the implantation parameters, we performed Stopping and Range of Ions in Matter (SRIM) simulations, a foundational and widely-used tool in ion-solid interaction physics[44]. This allowed us to calculate the Displacements Per Atom (DPA), a key metric quantifying the average number of atomic displacements[45]. As shown in Fig. 3c and 3e, DPA rate (DPA/s) was used to quantify the intensity of energy deposition. The simulation (Fig. 3c) reveals that ion energy profoundly dictates the resulting damage profile, with higher energies inducing more significant displacement cascades.

This simulation underpins our interpretation of the electrochemical data in Fig. 3d. This experiment was designed as a crucial control study: the total implantation dose was held constant at $1 \times 10^{16}$ ions/$cm^2$ for all samples. This normalization of the mass effects (i.e., the total number of implanted Pt atoms is nominally identical) allows us to isolate the contribution of the energy effects (defects and phase transitions) to the catalytic activity. The result is unequivocal: catalysts prepared at 10 keV and 20 keV exhibit performance vastly superior to that of the 5 keV sample. This suggests that higher implantation energies generate a sufficiently high density of modified sites to saturate the catalytic improvement, whereas the 5 keV energy remains insufficient to fully activate the support. This directly confirms that the energy-driven structural modification is a prerequisite for achieving high catalytic performance.

Extending this analysis to the broader SACs library highlights the complex interplay between the energy effects and the mass effects. Fig. 3e enables a clear deconvolution of these contributions. First, for lighter ions (Fe, Co, Ni, Cu), the relatively low DPA/s values indicate that the energy effects are insufficient to induce adequate structural reconstruction, resulting in generally modest performance where minor variations arise primarily from intrinsic chemical trends. In contrast, heavier ions (Pt, Ru, Mo) all generate significantly higher DPA/s values, satisfying the structural prerequisite for site generation. However, their catalytic outcomes diverge sharply: while Pt/MoS$_2$ delivers exceptional HER activity, Ru and Mo counterparts perform poorly. This stark contrast serves as a crucial proof that strong energy effects are a necessary but insufficient condition; the intrinsic activity of the metal species remains the decisive factor[46]. Finally, the superior performance of Pt/MoS$_2$ compared to Pt/Carbon underscores both the intrinsic differences in Pt anchoring capability and the support's response to irradiation. On one hand, MoS$_2$ is inherently a more favorable substrate for stabilizing Pt single atoms; on the other hand, its lattice is more susceptible to beneficial restructuring than radiation-hard carbon. This is further corroborated by ICP measurements, which reveal that the Pt loading on MoS$_2$ is approximately one order of magnitude higher than that on graphene under identical ion doses (Supplementary Fig. 11).

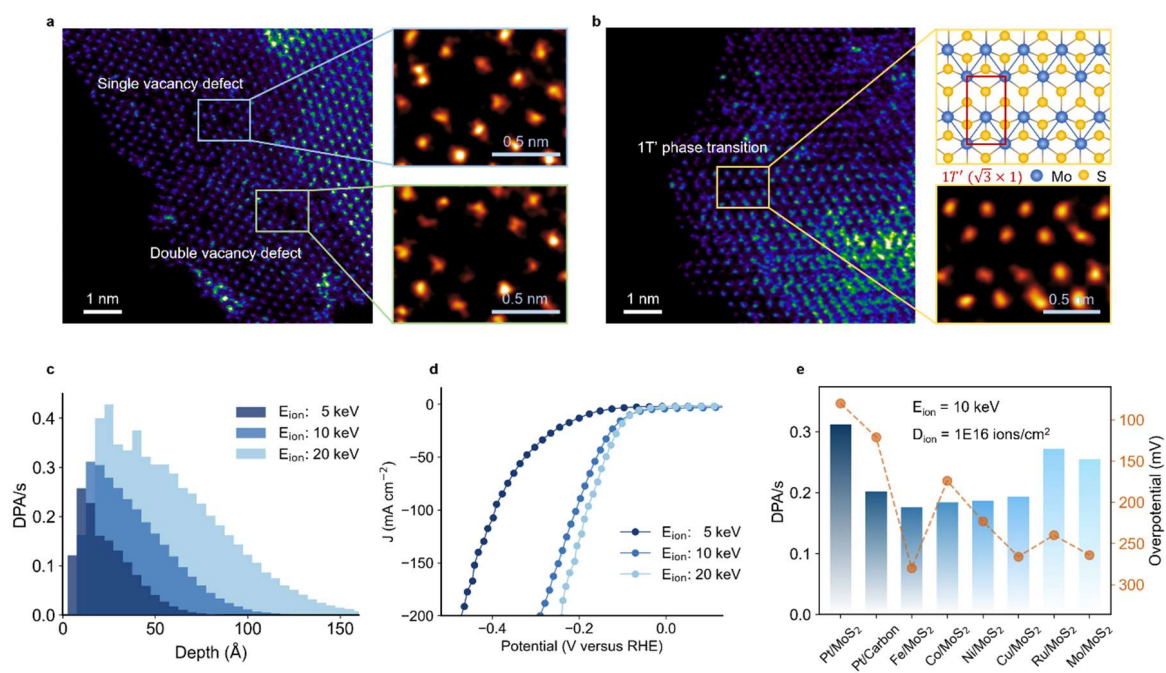

**Fig. 3. Synergistic Mechanism of Ion Implantation. a, b,** HAADF-STEM images of structural modifications induced by the energy effects: monovacancies and divacancies (a); local 1T' phase $MoS_2$ (b). **c**, SRIM simulation of DPA/s versus depth for 5, 10, and 20 keV Pt ion energies. **d**, HER polarization curves for catalysts prepared at constant dose ($1\times10^{16}$ ions/cm$^2$) but varying ion energies. **e**, Comparison of simulated DPA/s (bars) and measured HER overpotential (line) for typical SACs on $MoS_2$ and Carbon supports.

**Electrocatalytic performance and methodological advantages**

Benefiting from the synergistic mechanism and non-equilibrium nature discussed above, the ion implantation strategy effectively circumvents thermodynamic aggregation, thereby offering superior controllability over the synthesis of high-quality SACs. As a proof of concept, we evaluated the Pt/MoS$_2$ series as advanced electrocatalysts for HER. The HER performance of Pt/MoS$_2$ electrocatalysts prepared with different ion doses, where a unit dose (1.0 Pt/MoS$_2$) corresponds to $1 \times 10^{16}$ ions cm$^{-2}$, was measured in 1 M KOH solution using commercial 20wt% Pt/C as a reference (Fig. 4a). The 5.0 Pt/MoS$_2$ exhibits the highest HER activity, with an overpotential of only 26 mV to deliver the current density of 10 mA cm$^{-2}$. This performance is vastly superior to commercial 20wt% Pt/C (44 mV at 10 mA cm$^{-2}$) and pristine MoS$_2$ (266 mV at

10 mA cm$^{-2}$). The kinetic advantage is further confirmed by the Tafel plot, where 5.0 Pt/MoS$_2$ achieves the minimum slope of 46.8 mV dec$^{-1}$ ([Fig. 4b](#)), indicating the most rapid reaction kinetics among all control samples. Electrochemical impedance spectroscopy (EIS) further confirms the improved charge transfer efficiency, as 5.0 Pt/MoS$_2$ exhibits the lowest charge transfer resistances ($R_{ct}$) ([Supplementary Fig.12](#)). Electrochemically active surface area (ECSA), estimated from the double layer capacitance ($C_{dl}$), reveals that 5.0 Pt/MoS$_2$ possesses the largest value of 16.38 mF cm$^{-2}$ ([Supplementary Fig. 13](#)). This result confirms that 5.0 Pt/MoS$_2$ provides a higher density of accessible active sites, thereby contributing to its superior HER performance. In terms of mass activity, 5.0 Pt/MoS$_2$ delivers 864.23 A g Pt$^{-1}$ at an overpotential of 100 mV, which is 1.70, 1.52, 2.38 times higher than 0.1 Pt/MoS$_2$, 1 Pt/MoS$_2$, and 20wt% Pt/C, respectively ([Supplementary Fig. 14](#)). A benchmark comparison of the overpotential of 5.0 Pt/MoS$_2$ outperforms the majority of reported Pt-based alkaline electrocatalysts ([Fig. 4c](#) and [Supplementary Table 2)](#).

We further explored the quantitative correlation between synthesis parameters and performance to validate the controllability of this strategy. As shown in [Fig. 4d](#), the ICP-measured Pt content exhibits a strictly linear relationship with the implantation dose at lower levels. At higher doses, a saturation behavior is observed, likely due to the support loss caused by continuous energy deposition. Crucially, the catalytic overpotential displays a strong negative linear correlation ($R^2$ = 0.99) with the Pt loading ([Fig. 4e](#)). This behavior suggests that the non-equilibrium synthesis effectively decouples metal loading from cluster formation; unlike thermal processes where high loading inevitably leads to aggregation, our strategy maintains a high purity of SACs even at elevated implantation doses, thereby ensuring that performance scales linearly with active site density.

Furthermore, the catalyst exhibits exceptional stability. As shown in [Supplementary Fig. 15](#), the polarization curve exhibits no obvious degradation even after 10000 cyclic voltammetry cycles. A 100 h chronoamperometric test further confirms the excellent durability, with the overpotential remaining essentially unchanged ([Fig. 4f](#)). Post-stability characterizations, including transmission electron microscopy (TEM) and X-ray absorption spectroscopy (XAS) analyses (Supplementary Figs. 16-17), confirm that the overall morphology is well preserved and Pt remains atomically dispersed. This highlights the strong anchoring effect of the "energy-engineered" defects in stabilizing isolated atoms under harsh conditions.

The industrial potential is further underscored by the wafer-scale scalability. As shown in Supplementary Fig. 18, we evaluated the HER activity of 8 randomly selected regions from an 8-inch wafer. The measured overpotentials yield a mean value of 29.76 mV with a standard deviation of 4.20 mV, corresponding to a coefficient of variation (C. V.) of 14.1%. This low variance confirms the macroscopic uniformity of the synthesis. To demonstrate this practical viability, we constructed an electrolytic cell based on an anion exchange membrane (AEM) with 5.0 Pt/MoS$_2$ as the cathode catalyst and commercial iridium-coated nickel felt as the anode. The AEM electrolytic cell was constructed into a sandwich configuration with an area of 4 cm$^2$ (Supplementary Fig. 19). The device delivers 500 mA cm$^{-2}$ at a low voltage of 1.56 V (60 °C) and operates stably for 100 hours (Figs. 4g-h).

Beyond the model Pt/MoS$_2$ system, this strategy demonstrates exceptional universality and scalability. We evaluated the library of 36 SACs derived from the combinations illustrated in Fig. 1a, including complex multi-metal SACs, as evidenced by STEM and EDX mapping (Supplementary Figs. 20b and 20c). The electrocatalytic performance was summarized in the heat map (Supplementary Fig. 20a). Collectively, these results demonstrate that our ion implantation strategy offers a unique combination of high catalyst quality, precise controllability, macroscopic uniformity, and exceptional stability. This establishes it not only as a robust route for practical applications but also as a powerful high-throughput platform for the broader exploration of SACs.

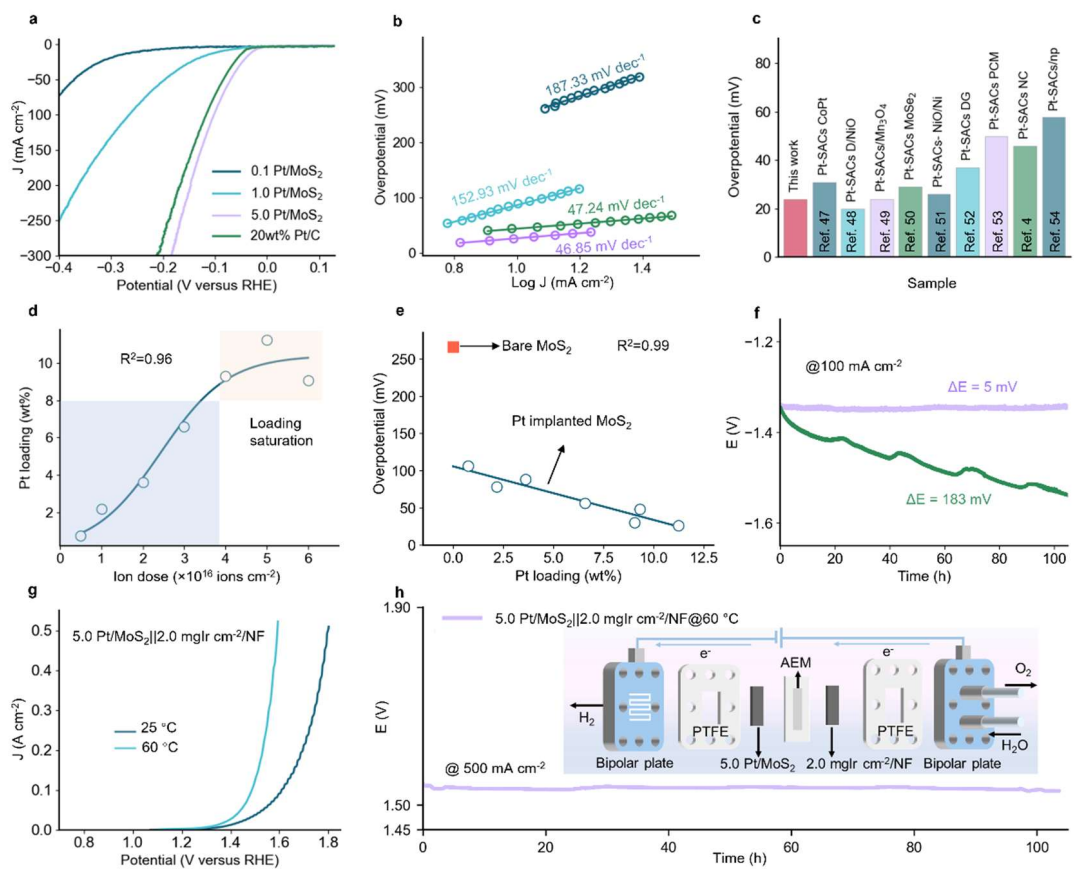

**Fig. 4**. **Measurements of electrocatalytic performance. a,** HER polarization curves at a scan rate of 5 mV/s and **b,** The Tafel slope of 0.1 Pt/MoS$_2$, 1.0 Pt/MoS$_2$, 5.0 Pt/MoS$_2$, 20wt% Pt/C. **c,** Comparison of the overpotential of the state of art Pt SACs at 10 mA cm$^{-2}$ in 1.0 M KOH[4,47-54]. **d,** Relationship between ion implantation dose and resulting metal loading in Pt/MoS$_2$. **e,** Relationship between the HER overpotential and Pt loading. **f,** Chronopotentiometry tests of 100 h in 1.0 M KOH at room temperature of 5.0 Pt/MoS$_2$ and 20wt% Pt/C. **g,** Linear sweep voltammetry (LSV) curves with a 4 cm$^2$ electrode area of 5.0 Pt/MoS$_2$/NF|| 2.0 mgIr cm$^{-2}$/NF for overall water splitting in AEM electrolyzer in 1.0 M KOH at room temperature and 60 °C, respectively. **h,** Chronopotentiometry test in 1.0 M KOH at 60 °C of 5.0 Pt/MoS$_2$/NF|| 2.0 mgIr cm$^{-2}$/NF at 500 mA cm$^{-2}$.

**Conclusion**

In summary, this work establishes ion implantation as a non-equilibrium paradigm for the synthesis of SACs. By exploiting the synergistic energy-mass effects inherent to energetic ion beams, metal atoms can be simultaneously introduced and stabilized through in situ defect and phase engineering, enabling precise atomic dispersion across diverse supports. Unlike conventional thermodynamic routes, this kinetically controlled process decouples metal loading from aggregation, allowing SACs quality and catalytic performance to scale linearly with active site density. The industrially mature, wafer-scale implantation capability further ensures high throughput, macroscopic uniformity, and quantitative controllability, bridging the gap between fundamental SACs design and practical manufacturing. As a representative example, Pt/$MoS_2$ exhibits outstanding hydrogen evolution activity and long-term stability, highlighting the effectiveness of this strategy in producing high-performance catalysts. More broadly, ion implantation offers a general and extensible platform for the rational design and high-throughput exploration of SACs, opening new opportunities for energy conversion and beyond.

**Acknowledgments**

This work was financially supported by the National Natural Science Foundation of China (Grant Nos. 52272033, 22494641, 52021006 and 52402041), the Beijing National Laboratory for Molecular Sciences (BNLMS-CXTD-202001) and the Shenzhen Science and Technology Innovation Commission (KQTD20221101115627004).

**Reference**

1       Qiao, B. *et al.* Single-atom catalysis of CO oxidation using Pt1/FeOx. *Nature Chemistry* **3**, 634-641, doi:10.1038/nchem.1095 (2011).

2       Wang, A., Li, J. & Zhang, T. Heterogeneous single-atom catalysis. *Nature Reviews Chemistry* **2**, 65-81, doi:10.1038/s41570-018-0010-1 (2018).

3       Cao, L. *et al.* Identification of single-atom active sites in carbon-based cobalt catalysts during electrocatalytic hydrogen evolution. *Nature Catalysis* **2**, 134-141, doi:10.1038/s41929-018-0203-5 (2019).

4       Fang, S. *et al.* Uncovering near-free platinum single-atom dynamics during electrochemical hydrogen evolution reaction. *Nature Communications* **11**, 1029, doi:10.1038/s41467-020-14848-2 (2020).

5       Fang, G. *et al.* Retrofitting Zr-Oxo Nodes of UiO-66 by Ru Single Atoms to Boost Methane Hydroxylation with Nearly Total Selectivity. *Journal of the American Chemical Society* **145**, 13169-13180, doi:10.1021/jacs.3c02121 (2023).

6       Wang, Z. *et al.* Aunano-Fe1 tandem catalysis for promoted methane conversion to acetic acid by O2 oxidation. *Applied Catalysis B: Environment and Energy* **378**, 125632, doi:https://doi.org/10.1016/j.apcatb.2025.125632 (2025).

7       Chang, J. *et al.* Synthesis of ultrahigh-metal-density single-atom catalysts via metal sulfide-mediated atomic trapping. *Nature Synthesis* **3**, 1427-1438, doi:10.1038/s44160-024-00607-4 (2024).

8       Hai, X. *et al.* Scalable two-step annealing method for preparing ultra-high-density single-atom catalyst libraries. *Nature Nanotechnology* **17**, 174-181, doi:10.1038/s41565-021-01022-y (2022).

9       Kaushik, S. *et al.* Universal Synthesis of Single-Atom Catalysts by Direct Thermal Decomposition of Molten Salts for Boosting Acidic Water Splitting. *Advanced Materials* **36**, 2401163, doi:https://doi.org/10.1002/adma.202401163 (2024).

10      Tian, L. *et al.* Precise arrangement of metal atoms at the interface by a thermal printing strategy. *Proc Natl Acad Sci U S A* **120**, e2310916120, doi:10.1073/pnas.2310916120 (2023).

11      Yang, H. *et al.* A universal ligand mediated method for large scale synthesis of transition metal single atom catalysts. *Nature Communications* **10**, 4585, doi:10.1038/s41467-019-12510-0 (2019).

12  Liu, S. *et al.* Chemical Vapor Deposition for Atomically Dispersed and Nitrogen Coordinated Single Metal Site Catalysts. *Angewandte Chemie International Edition* **59**, 21698-21705, doi:https://doi.org/10.1002/anie.202009331 (2020).

13  Fonseca, J. & Lu, J. Single-Atom Catalysts Designed and Prepared by the Atomic Layer Deposition Technique. *ACS Catalysis* **11**, 7018-7059, doi:10.1021/acscatal.1c01200 (2021).

14  Datye, A. & Wang, Y. Atom trapping: a novel approach to generate thermally stable and regenerable single-atom catalysts. *National Science Review* **5**, 630-632, doi:10.1093/nsr/nwy093 (2018).

15  Hansen, T. W., DeLaRiva, A. T., Challa, S. R. & Datye, A. K. Sintering of Catalytic Nanoparticles: Particle Migration or Ostwald Ripening? *Accounts of Chemical Research* **46**, 1720-1730, doi:10.1021/ar3002427 (2013).

16  Liu, L. *et al.* Dealuminated Beta zeolite reverses Ostwald ripening for durable copper nanoparticle catalysts. *Science* **383**, 94-101, doi:10.1126/science.adj1962 (2024).

17  Su, Y.-Q. *et al.* Stability of heterogeneous single-atom catalysts: a scaling law mapping thermodynamics to kinetics. *npj Computational Materials* **6**, 144, doi:10.1038/s41524-020-00411-6 (2020).

18  Du, K. *et al.* Mixed-valence palladium single-atom catalyst induced by hybrid TiO2-graphene through a photochemical strategy. *Applied Surface Science* **625**, 157115, doi:https://doi.org/10.1016/j.apsusc.2023.157115 (2023).

19  Fu, W. *et al.* Photoinduced loading of electron-rich Cu single atoms by moderate coordination for hydrogen evolution. *Nature Communications* **13**, 5496, doi:10.1038/s41467-022-33275-z (2022).

20  Shen, S., Zhao, L. & Zhang, J. Promising approach for preparing metallic single-atom catalysts: electrochemical deposition. *Frontiers in Energy* **16**, 537-541, doi:10.1007/s11708-022-0837-5 (2022).

21  Zhang, Z. *et al.* Electrochemical deposition as a universal route for fabricating

single-atom catalysts. *Nature Communications* **11**, 1215, doi:10.1038/s41467-020-14917-6 (2020).

22  Liu, J.-C., Tang, Y., Wang, Y.-G., Zhang, T. & Li, J. Theoretical understanding of the stability of single-atom catalysts. *National Science Review* **5**, 638-641, doi:10.1093/nsr/nwy094 (2018).

23  Zhang, J. *et al.* Cation vacancy stabilization of single-atomic-site Pt1/Ni(OH)x catalyst for diboration of alkynes and alkenes. *Nature Communications* **9**, 1002, doi:10.1038/s41467-018-03380-z (2018).

24  Pan, Y. *et al.* Design of Single-Atom Co–N5 Catalytic Site: A Robust Electrocatalyst for CO2 Reduction with Nearly 100% CO Selectivity and Remarkable Stability. *Journal of the American Chemical Society* **140**, 4218-4221, doi:10.1021/jacs.8b00814 (2018).

25  Wang, L. *et al.* Boosting Activity and Stability of Metal Single-Atom Catalysts via Regulation of Coordination Number and Local Composition. *Journal of the American Chemical Society* **143**, 18854-18858, doi:10.1021/jacs.1c09498 (2021).

26  Li, Z. & Chen, F. Ion beam modification of two-dimensional materials: Characterization, properties, and applications. *Applied Physics Reviews* **4**, 011103, doi:10.1063/1.4977087 (2017).

27  Wang, G. *et al.* Synthesis of Layer-Tunable Graphene: A Combined Kinetic Implantation and Thermal Ejection Approach. *Advanced Functional Materials* **25**, 3666-3675, doi:https://doi.org/10.1002/adfm.201500981 (2015).

28  Xie, Y. *et al.* Nano-seeding catalysts for high-density arrays of horizontally aligned carbon nanotubes with wafer-scale uniformity. *Nature Communications* **16**, 149, doi:10.1038/s41467-024-55515-0 (2025).

29  Maharana, B. *et al.* Defect-engineered MnO2 nanoparticles by low-energy ion beam irradiation for enhanced electrochemical energy storage applications. *Electrochimica Acta* **464**, 142868, doi:https://doi.org/10.1016/j.electacta.2023.142868 (2023).

30  Zhou, X. *et al.* TiO2 Nanotubes: Nitrogen-Ion Implantation at Low Dose Provides Noble-Metal-Free Photocatalytic H2 -Evolution Activity. *Angew Chem Int Ed Engl* **55**,

3763-3767, doi:10.1002/anie.201511580 (2016).

31  Li, Y. *et al.* Highly-defective graphene as a metal-free catalyst for chemical vapor deposition growth of graphene glass. *Carbon* **187**, 272-279, doi:https://doi.org/10.1016/j.carbon.2021.11.019 (2022).

32  Chen, Y. *et al.* Tuning Electronic Structure of Single Layer MoS2 through Defect and Interface Engineering. *ACS Nano* **12**, 2569-2579, doi:10.1021/acsnano.7b08418 (2018).

33  He, Z. *et al.* Defect Engineering in Single-Layer MoS2 Using Heavy Ion Irradiation. *ACS Applied Materials & Interfaces* **10**, 42524-42533, doi:10.1021/acsami.8b17145 (2018).

34  Li, Y. *et al.*

35  Shi, Z. *et al.* Phase-dependent growth of Pt on MoS2 for highly efficient H2 evolution. *Nature* **621**, 300-305, doi:10.1038/s41586-023-06339-3 (2023).

36  Chen, Y. *et al.* Isolated Single Iron Atoms Anchored on N-Doped Porous Carbon as an Efficient Electrocatalyst for the Oxygen Reduction Reaction. *Angewandte Chemie International Edition* **56**, 6937-6941, doi:https://doi.org/10.1002/anie.201702473 (2017).

37  Gu, J., Hsu, C.-S., Bai, L., Chen, H. M. & Hu, X. Atomically dispersed $Fe^{3+}$ sites catalyze efficient CO2 electroreduction to CO. *Science* **364**, 1091-1094, doi:10.1126/science.aaw7515 (2019).

38  Chakraborty, A., González Hernández, R., Šmejkal, L. & Sinova, J. Strain-induced phase transition from antiferromagnet to altermagnet. *Physical Review B* **109**, 144421, doi:10.1103/PhysRevB.109.144421 (2024).

39  Lin, Y.-C., Dumcenco, D. O., Huang, Y.-S. & Suenaga, K. Atomic mechanism of the semiconducting-to-metallic phase transition in single-layered MoS2. *Nature Nanotechnology* **9**, 391-396, doi:10.1038/nnano.2014.64 (2014).

40  Wang, D. *et al.* Phase engineering of a multiphasic 1T/2H MoS2 catalyst for highly efficient hydrogen evolution. *Journal of Materials Chemistry A* **5**, 2681-2688, doi:10.1039/C6TA09409K (2017).

41  Li, Y. *et al.* Synergistic Pt doping and phase conversion engineering in two-dimensional MoS2 for efficient hydrogen evolution. *Nano Energy* **84**, 105898, doi:https://doi.org/10.1016/j.nanoen.2021.105898 (2021).

42  Lukowski, M. A. *et al.* Enhanced Hydrogen Evolution Catalysis from Chemically Exfoliated Metallic MoS2 Nanosheets. *Journal of the American Chemical Society* **135**, 10274-10277, doi:10.1021/ja404523s (2013).

43  Wang, L. *et al.* Self-Optimization of the Active Site of Molybdenum Disulfide by an Irreversible Phase Transition during Photocatalytic Hydrogen Evolution. *Angewandte Chemie International Edition* **56**, 7610-7614, doi:https://doi.org/10.1002/anie.201703066 (2017).

44  Ziegler, J. F. & Biersack, J. P. in *Treatise on Heavy-Ion Science: Volume 6: Astrophysics, Chemistry, and Condensed Matter*   (ed D. Allan Bromley)  93-129 (Springer US, 1985).

45  Piñera Hernández, I., Cruz Inclán, C. M., Leyva Fabelo, A. & Abreu Alfonso, Y. Calculation of displacements per atom distributions in solid materials. *Nucleus*, 39-44 (2007).

46  Nørskov, J. K. *et al.* Trends in the Exchange Current for Hydrogen Evolution. *Journal of The Electrochemical Society* **152**, J23, doi:10.1149/1.1856988 (2005).

47  Yang, W. *et al.* Tuning the Cobalt–Platinum Alloy Regulating Single-Atom Platinum for Highly Efficient Hydrogen Evolution Reaction. *Advanced Functional Materials* **32**, 2205920, doi:https://doi.org/10.1002/adfm.202205920 (2022).

48  Yan, Y. *et al.* Atomic-Level Platinum Filling into Ni-Vacancies of Dual-Deficient NiO for Boosting Electrocatalytic Hydrogen Evolution. *Advanced Energy Materials* **12**, 2200434, doi:https://doi.org/10.1002/aenm.202200434 (2022).

49  Wei, J. *et al.* In situ precise anchoring of Pt single atoms in spinel Mn3O4 for a highly efficient hydrogen evolution reaction. *Energy & Environmental Science* **15**, 4592-4600, doi:10.1039/D2EE02151J (2022).

50  Shi, Y. *et al.* Electronic metal–support interaction modulates single-atom platinum catalysis for hydrogen evolution reaction. *Nature Communications* **12**, 3021, doi:10.1038/s41467-021-23306-6 (2021).

51  Zhou, K. L. *et al.* Platinum single-atom catalyst coupled with transition metal/metal oxide heterostructure for accelerating alkaline hydrogen evolution reaction. *Nature Communications* **12**, 3783, doi:10.1038/s41467-021-24079-8 (2021).

52  Yang, Q. *et al.* Single Carbon Vacancy Traps Atomic Platinum for Hydrogen Evolution Catalysis. *Journal of the American Chemical Society* **144**, 2171-2178, doi:10.1021/jacs.1c10814 (2022).

53  Zhang, H. *et al.* Dynamic traction of lattice-confined platinum atoms into mesoporous carbon matrix for hydrogen evolution reaction. *Science Advances* **4**, eaao6657, doi:doi:10.1126/sciadv.aao6657 (2018).

54  Jiang, K. *et al.* Single platinum atoms embedded in nanoporous cobalt selenide as electrocatalyst for accelerating hydrogen evolution reaction. *Nature Communications* **10**, 1743, doi:10.1038/s41467-019-09765-y (2019).

# Supplementary Information

# A non-equilibrium strategy for the general synthesis of single-atom catalysts

Yue Li,[1,2,§] Yang Xu,[1,§] Yunbiao Zhao,[1] Mingwei Cui,[2] Xiner Chen,[1] Liu Qian,[2] Jin Zhang,[2] Xueting Feng,[3]* Ziqiang Zhao[1,2]*

[1] State Key Laboratory of Nuclear Physics and Technology, School of Physics, Peking University, Beijing, 100871, China.

[2] School of Materials Science and Engineering, Peking University, Beijing, 100871, China.

[3] Department of Chemistry, Capital Normal University, Beijing, 100048, China.

[§] These authors contributed equally to this work: Yue Li and Yang Xu

* E-mail: zqzhao@pku.edu.cn; fengxt@cnu.edu.cn

**Materials**

The materials used in the experiment, including thin-layer $MoS_2$ with small flake diameter, multi-walled carbon nanotubes (short), purity >95%, length 0.5-2 μm, diameter <8 nm, and reagent grade graphene with thickness of 1-3 nm, number of layers <3, and size >50 μm, all of which are purchased from XFNANO. Graphdiyne was obtained by the reported work[1], 20 wt% Pt/C was obtained from MACKLIN, dicalcium phosphate (KCl, solid, >99.5%), potassium hydroxide (KOH, solid, >85%), Isopropyl alcohol (($CH_3$)$_2$CHOH, solution, >99.7%) were obtained from TGREAG. Relative hydrophilic carbon paper was obtained from SINERO. Nafion® D-520 dispersion, 5%w/w in water and 1-propanol, >1meq/g exchange capacity, was purchased from GZAMHOO.

**Catalyst preparation**

Firstly, the carbon paper was cut into small pieces of 1×1.5 cm$^2$, and then sonicated with acetone, ethanol, and deionized water to remove surface oils and impurities, respectively, and baked in a vacuum drying oven at 80 °C for 12 h. KOH solids weighed 28.05 g was dissolved in 500 ml of deionized water, ultrasonication to make it fully dissolved in a volumetric flask to form 1.0 M/L KOH, prior to each experiment, argon was bubbled through the solution to remove dissolved oxygen.

In a typical synthesis, 12 mg $MoS_2$ was dispersed into 1100 μl of isopropanol and 100 μl of Nafion mixture, sonicated for 30 min and then 100 μl of the mixture was drop-coated onto 1×1 cm$^2$ carbon paper, after which it was put in a vacuum drying oven at 80 °C for 12 h to dry and set aside. 6 mg each of graphene, multi-walled carbon nanotubes, and graphdiyne were separately dispersed in a mixed solution containing 1100 μl of isopropanol and 100 μl of Nafion. A 200 μL aliquot was taken from each dispersion for drop-coating onto the electrode surface. The remaining dispersion for each material was treated in the same manner as that for molybdenum disulfide. Then, the collector coated with $MoS_2$, graphene, etc. were taken for ion implantation to load the catalysts, including Fe, Co, Ni, Cu, Pt, and so on. The energy varied from 5 keV to 20 keV and the doses ranged from 1×10$^{16}$ ions cm$^{-2}$ to 6×10$^{16}$ ions cm$^{-2}$.

**Physicochemical characterization.**

X-ray photoelectron spectroscopy (XPS) was carried out on an AXIS Ultra photoelectron spectrometer using monochromatic Al Kα radiation (Kratos Analytical). Transmission electron microscopy (TEM) and high-resolution TEM images were recorded using a FEI Tecnai F20 operating at 200 kV. Aberration-corrected high-angle annular dark-field scanning transmission electron microscopy (HAADF-STEM) images were acquired on JEOL ARM 300F operated at 300 kV equipped with a JEOL ETA corrector and an SDD-type EDX detector with the attainable energy resolution of 133 eV. The X-ray absorption fine structure (XAFS) measurements were carried out on the sample at 1W1B station in Beijing Synchrotron Radiation Facility (BSRF). The samples were detected in fluorescence mode with a Lytle detector, and the data processing of FT-EXAFS is carried out by ATHENA program. The inductively coupled plasma optical emission spectroscopy (ICP-OES, Prodigy 7) was used to analyze the Pt content.

**EXAFS fitting parameters**

The obtained XAFS data was processed in Athena (version 0.9.26) for background, pre-edge line and post-edge line calibrations. Then Fourier transformed fitting was carried out in Artemis (version 0.9.26). The $k^2$ weighting, k-range of 3-12.421 Å$^{-1}$ and R range of 1.5-3.5 Å were used for the fitting of Ru foil; Pt-1: k-range of 3-12.428 Å$^{-1}$ and R range of 1.3-2.5 Å; Pt-2：k-range of 2.6-13.412 Å$^{-1}$ and R range of 1.3-2.5 Å; Pt-3：k-range of 2.8-10.425 Å$^{-1}$ and R range of 1.2-2.7 Å were used for the fitting of samples. The four parameters, coordination number, bond length, Debye-Waller factor and $E_0$ shift (CN, R, $\Delta E_0$, $\sigma^2$) were fitted without anyone was fixed.

**Wavelet fitting parameters**

For Wavelet Transform analysis, the χ(k) exported from Athena was imported into the Hama Fortran code. The parameters were listed as follow: R range, 0-6 Å, k range, 0-15 Å$^{-1}$ for samples; k weight, 2; and Morlet function with κ=5, σ=2 was used as the mother wavelet to provide the overall distribution.

**Electrochemical Measurement**

The electrochemical measurements of the prepared catalysts were measured in the standard three-electrode test system on the CHI760E electrochemical workstation. The working electrode was carbon paper coated with the prepared catalyst with an exposed geometric surface area of 1 cm², the counter electrode was the Pt electrode, the reference electrode was the Ag/AgCl electrode under alkaline conditions. Cyclic voltammetry (CV), linear sweep voltammetry (LSV), electrochemical impedance spectroscopy (EIS), electrochemical active surface area (ECSA) and double-layer capacitance curve ($C_{dl}$) were used to evaluate the HER catalytic activity, all LSV curves were IR-corrected by 80%. All polarization curves were carried out in 1.0 M/L KOH solutions after a continuous cyclic voltammetry. The overpotential can be calculated according to the Nernst equation, the specific conversion formula with standard hydrogen electrode potential (RHE) was as follows: $E_{RHE} = E_{Ag/AgCl} + 0.0591 \times pH + E^*_{Ag/AgCl}(0.197V)$ (in alkaline media). As a control experiment, commercial Pt/C (20 wt% Pt loading, Johnson Matthey; 1.0 mg) was also placed as working electrode for measurements.

Prior to each measurement, research-grade Argon gas was bubbled through the 1.0 M/L KOH solution to remove dissolved $O_2$. LSV and CV measurements were made in the potential range [-0.9V, 0.2V] versus RHE with scanning rates of 5 and 100 mV/s, respectively. The EIS measurements were carried out in the same configuration at the potential of 10 mA/cm² from $10^{-1}$ to $10^{-5}$ Hz with a voltage of 5 mV. In order to correct the polarization data from dynamic voltage (IR) drop effects, the resistance values for the uncompensated solution were calculated from EIS measurements. The ECSA was measured by cyclic voltammetry scanning at different scanning rates (20-120 mV/s) in a non-faradic potential range and the double-layer capacitance curve $C_{dl}$ was obtained as the slope of electrochemically active area integrals at different sweep rates.

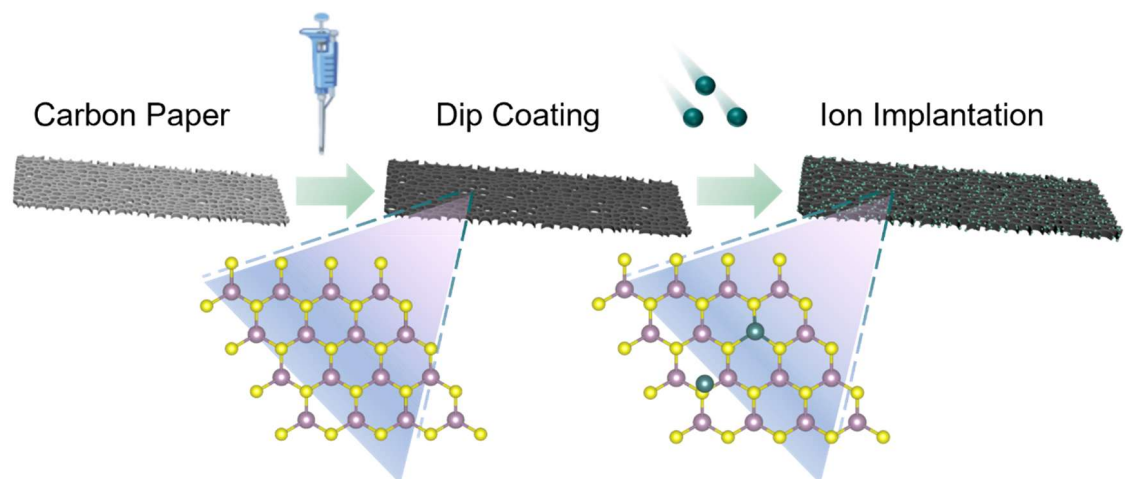

**Figure S1**. Schematic diagram of the ion implantation process for fabricating single-atom catalysts.

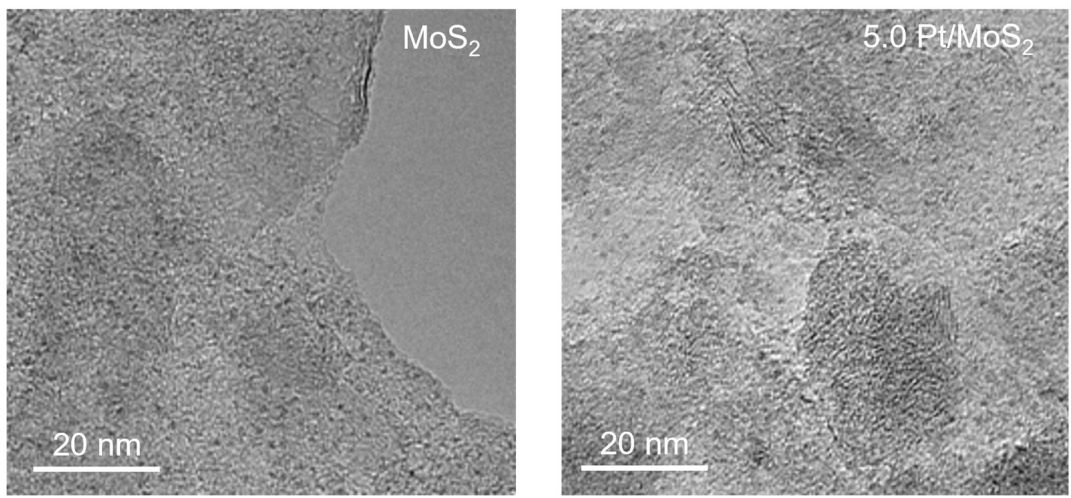

**Figure S2**. TEM images of (a) $MoS_2$ and (b) 5.0 $Pt/MoS_2$.

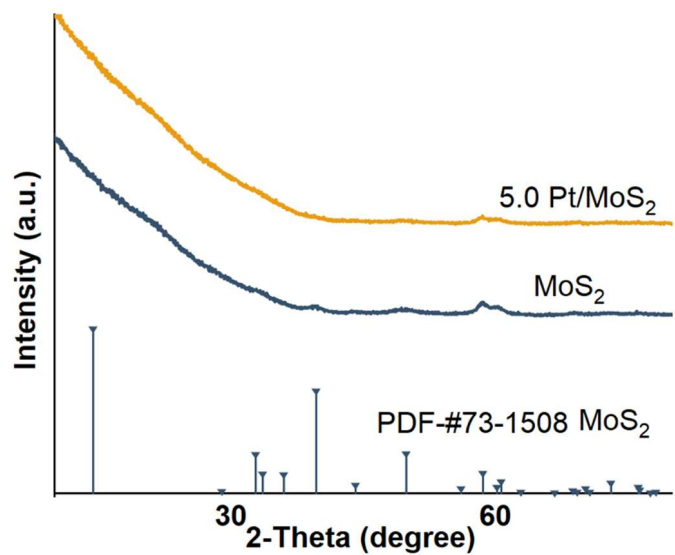

**Figure S3**. XRD patterns of $MoS_2$ and 5.0 Pt/$MoS_2$.

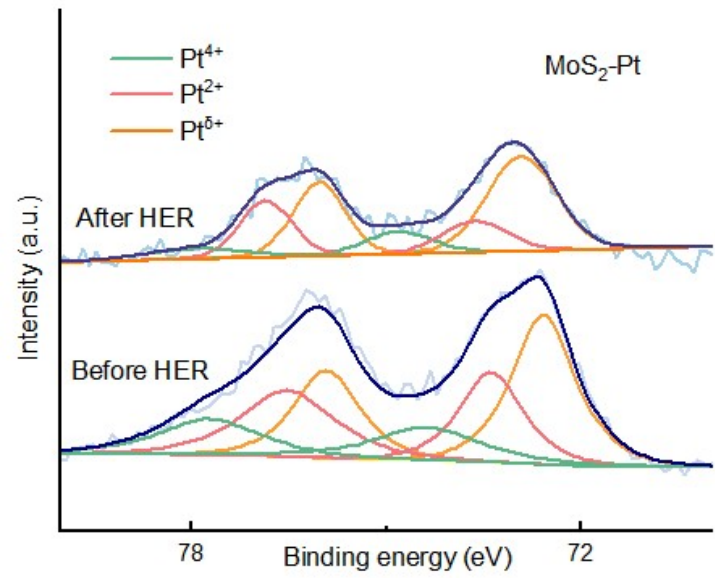

**Figure S4**. Pt 4f XPS spectra of 5.0 Pt/MoS$_2$.

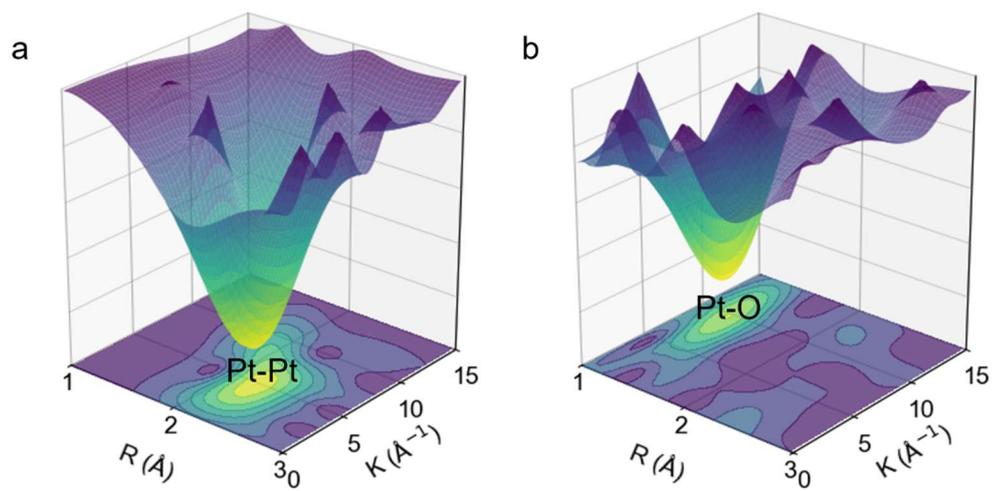

**Figure S5**. The 3D contour maps of (WT)-EXAFS spectra of **a,** Pt foil and **b,** PtO$_2$.

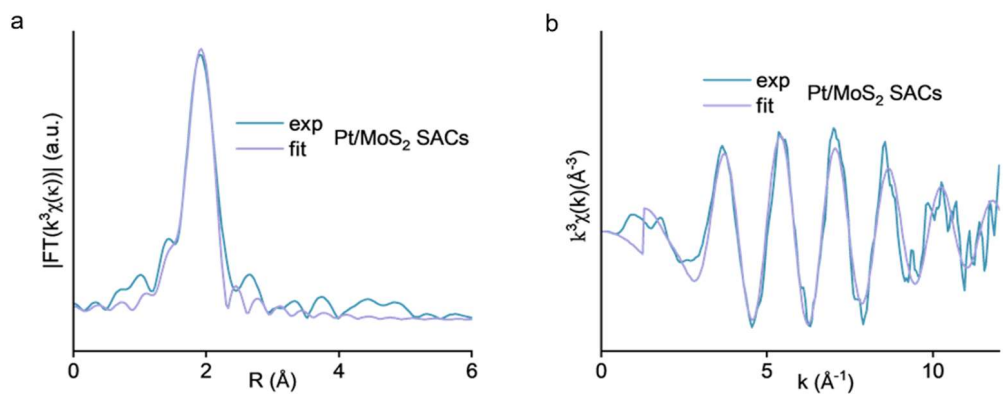

**Figure S6**. **a,** EXAFS fitting curves of 5.0 Pt/MoS$_2$ at the *R* space. **b,** EXAFS fitting curves of 5.0 Pt/MoS$_2$ at the *k* space.

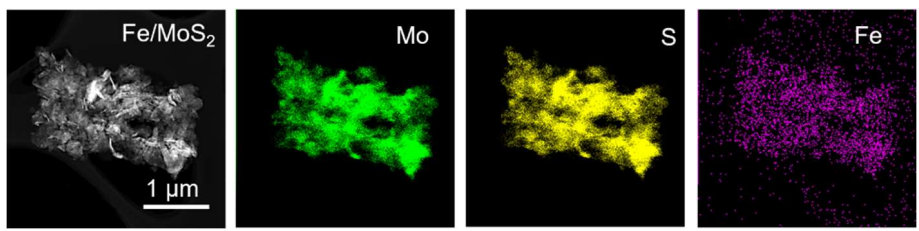

**Figure S7.** STEM image and corresponding EDX elemental mapping images of Fe/MoS$_2$.

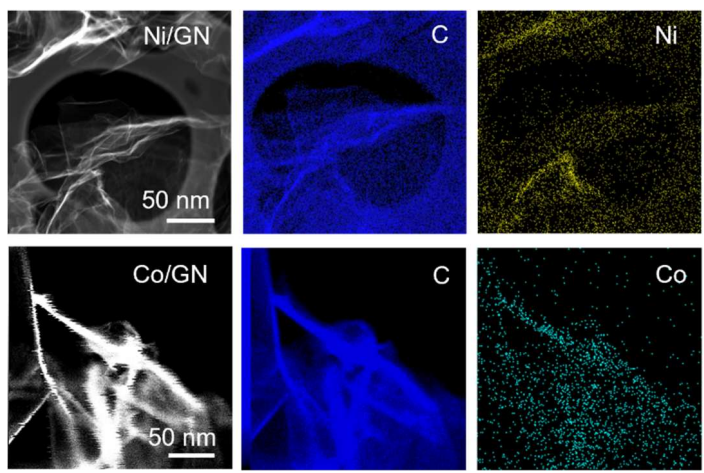

**Figure S8**. STEM images and corresponding EDX elemental mapping images of Ni/GN, and Co/GN.

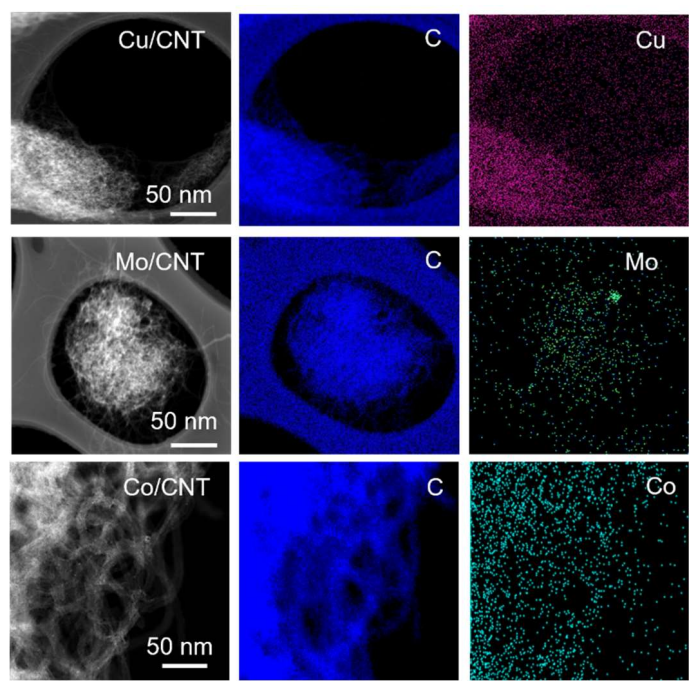

**Figure S9**. STEM images and corresponding EDX elemental mapping images of Cu/CNT, Mo/CNT, and Co/CNT.

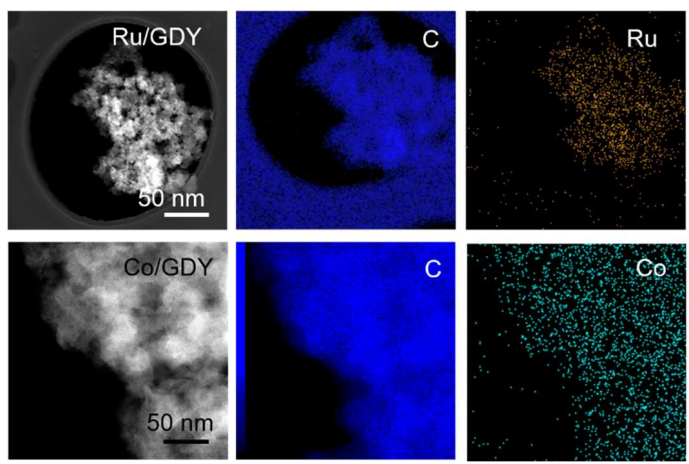

**Figure S10**. STEM images and corresponding EDX elemental mapping images of Ru/GDY, Co/GDY.

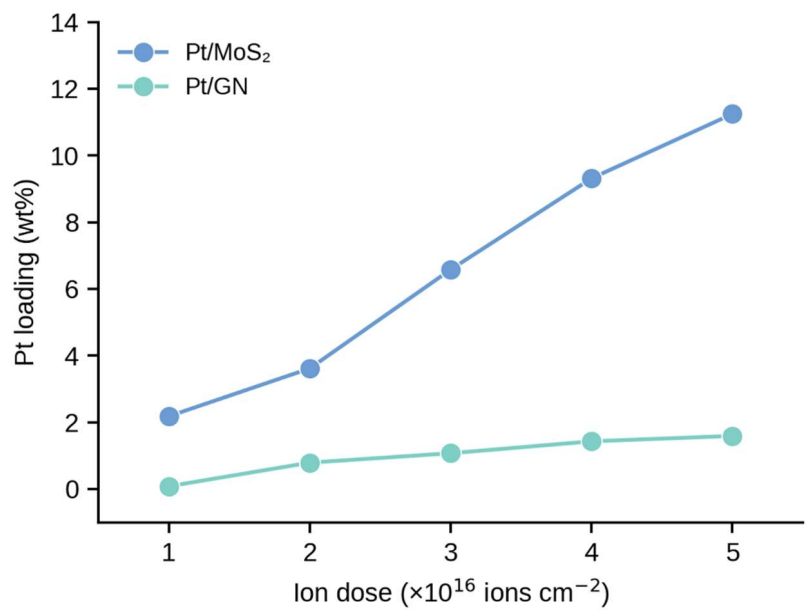

**Figure S11**. Comparison of the relationship between ion implantation dose and resulting metal loading in Pt/MoS$_2$ and Pt/GN.

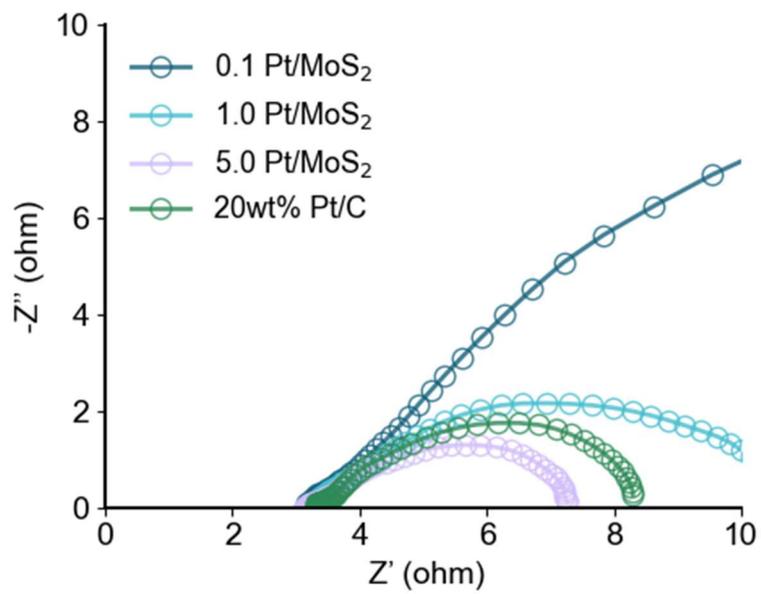

**Figure S12**. EIS plots of 0.1 Pt/MoS$_2$, 1.0 Pt/MoS$_2$, 5.0 Pt/MoS$_2$, 20wt% Pt/C.

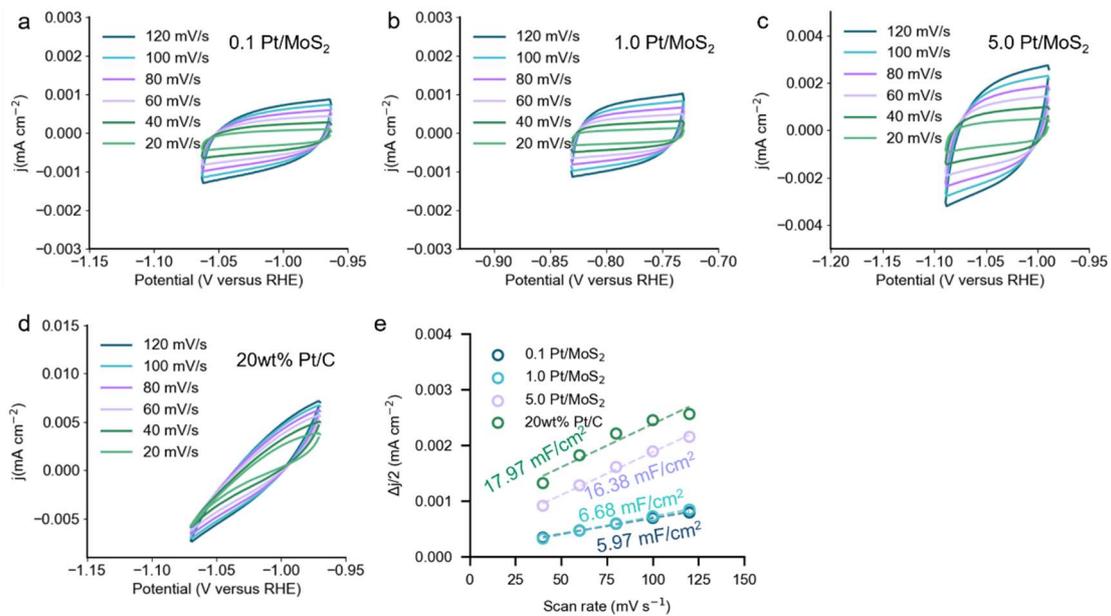

**Figure S13**. ECSA of **a,** 0.1 Pt/MoS$_2$, **b,** 1.0 Pt/MoS$_2$, **c,** 5.0 Pt/MoS$_2$, **d,** 20wt% Pt/C and **e,** C$_{dl}$ of 0.1 Pt/MoS$_2$, 1.0 Pt/MoS$_2$, 5.0 Pt/MoS$_2$, 20wt% Pt/C.

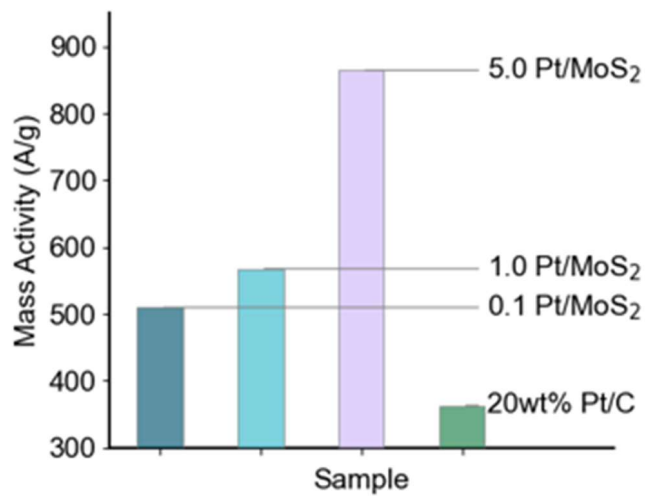

**Figure S14**. The mass activity at 100 mV of 0.1 Pt/MoS$_2$, 1.0 Pt/MoS$_2$, 5.0 Pt/MoS$_2$, 20wt% Pt/C.

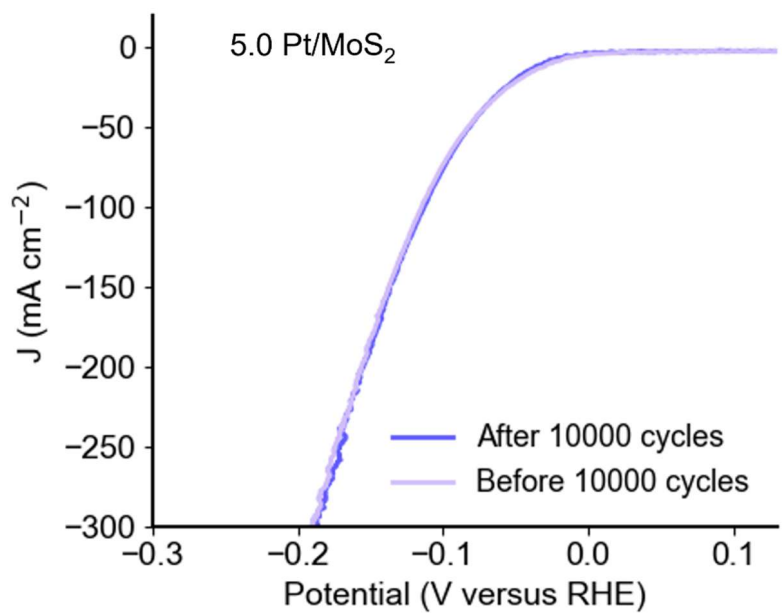

**Figure S15**. The LSV curves before and after 10000 cyclic voltammetry cycles.

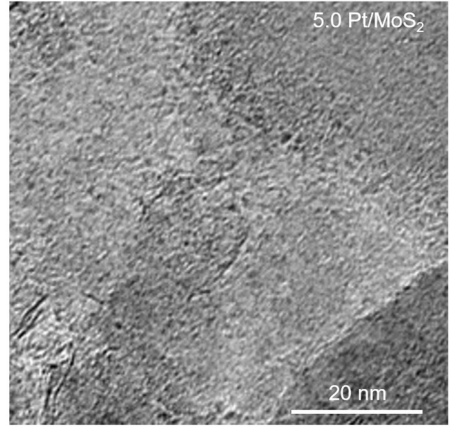

**Figure S16**. TEM image of 5.0 Pt/MoS$_2$ after 10000 cyclic voltammetry cycles.

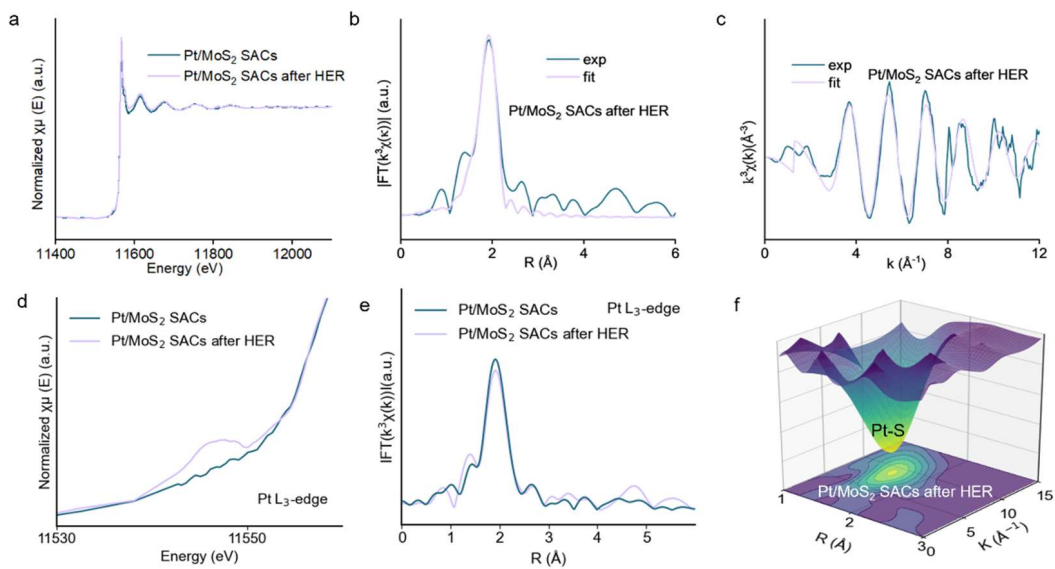

**Figure S17**. Structural characterization of Pt/MoS$_2$ SACs before and after HER by XANES. **a,** Normalized Pt L$_3$-edge XANES spectra of Pt/MoS$_2$ SACs before and after HER. **b,** FT-EXAFS at the Pt L$_3$-edge for Pt/MoS$_2$ SACs after HER, showing the experimental data and the fitting result. **c,** k$^3$-weighted χ(k) EXAFS oscillations in k-space for Pt/MoS$_2$ SACs after HER, with experimental data and the corresponding fit. **d,** Magnified view of the pre-edge region in (a) for detailed comparison. **e,** Comparison of the FT-EXAFS spectra for Pt/MoS$_2$ SACs before and after HER. **f,** WT-EXAFS signal for Pt/MoS$_2$ SACs after HER.

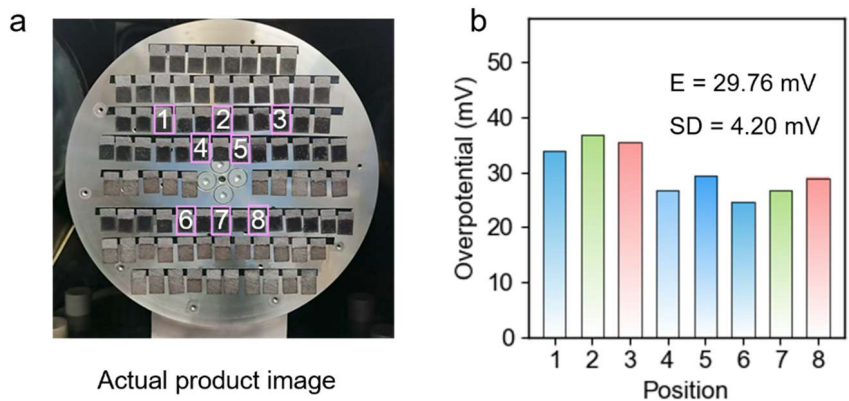

**Figure S18. a,** Schematic illustrating the random sampling of Pt/MoS$_2$ ion-implanted specimens from the fabrication batch. **b,** Electrochemical overpotential measurements obtained from the selected samples.

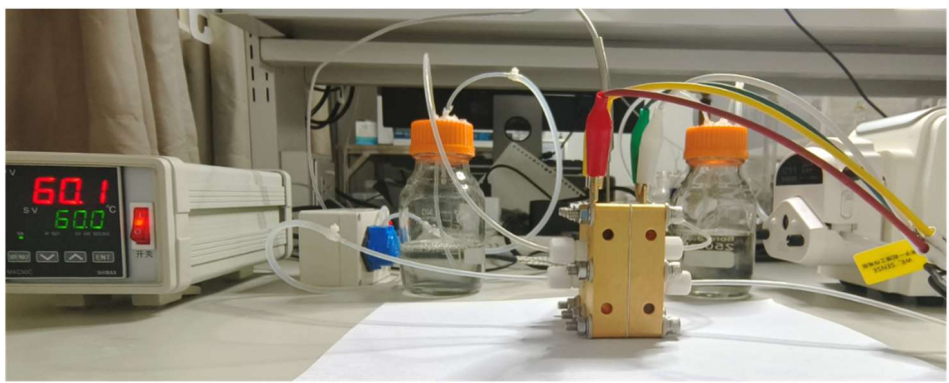

**Figure S19**. Anion exchange membrane (AEM) electrolyzer device.

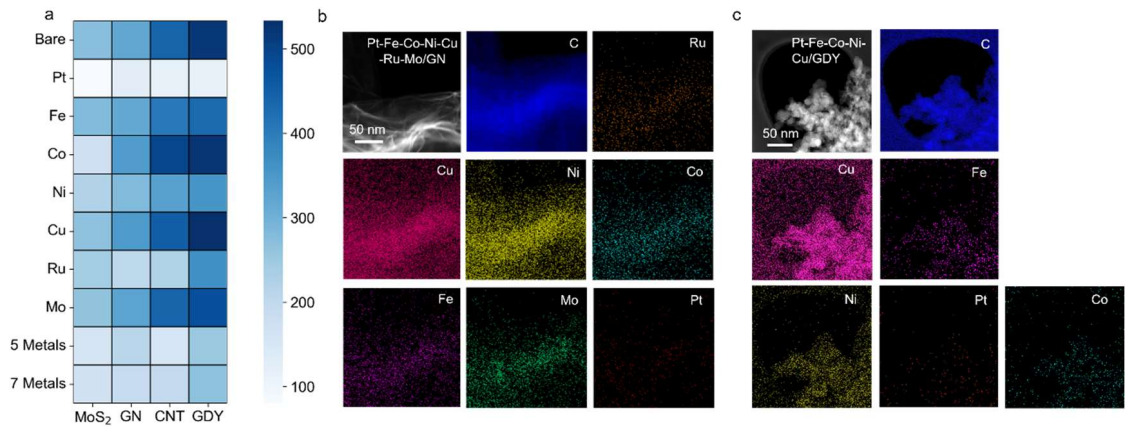

**Figure S20. a,** The overpotential values at 10 mA cm$^{-2}$ for different combinations of metals and substrates. STEM images and corresponding EDX elemental mapping images of **b,** Pt-Fe-Co-Ni-Cu-Ru-Mo/GN and **c,** Pt-Fe-Co-Ni-Cu/GDY.

**Table S1.** EXAFS fitting parameters at the Pt L3-edge for various samples ($S_0^2$=0.79).

| Sample | Shell | CN | R(Å) | $\sigma^2$(Å$^2$) | $\Delta E_0$ | R factor |
|---|---|---|---|---|---|---|
| Pt foil | Pt-Pt | 12 | 2.81±0.05 | 0.0042±0.0003 | 7.6±0.6 | 0.004 |
| Pt-1 | Pt-S | 5.2±0.8 | 2.42±0.09 | 0.0035±0.0020 | -6.0±1.8 | 0.020 |
| Pt-2 | Pt-S | 5.2±0.8 | 2.42±0.09 | 0.0044±0.0021 | -6.1±1.7 | 0.026 |
| Pt-3 | Pt-S | 4.2±0.7 | 2.42±0.10 | 0.0019±0.002 | -5.8±1.6 | 0.018 |

*N*: coordination numbers; *R*: bond distance; $\sigma^2$: Debye-Waller factors; $\Delta E_0$: the inner potential correction. *R* factor: goodness of fit.

Table S2. HER performance comparison with state-of-the-art catalysts.

| Catalyst | Overpotential(mV) | Tafel (mV dec$^{-1}$) | Reference |
|---|---|---|---|
| **Pt-SACs Co(OH)$_2$@Ag NWs [2]** | 29 | 35.72 | Energy Environ. Sci., 2020, 13,3082 |
| **Pt-SACs CoPt [3]** | 31 | 43.65 | Adv. Funct. Mater. 2022, 32, 2205920 |
| **Pt-SACs D NiO [4]** | 20 | 31.1 | Adv. Energy Mater. 2022, 12, 2200434 |
| **Ru-SACs ECM [5]** | 83 | 59 | Adv. Energy Mater. 2020, 10, 2000882 |
| **Pt-SACs/Mn$_3$O$_4$ [6]** | 24 | 54 | Energy Environ. Sci., 2022, 15, 4592 |
| **Ru-SACs MoO$_2$ [7]** | 29 | 44 | J. Mater. Chem. A,2017,5,5475–5485 |
| **Ni-SACs-MoS$_2$ [8]** | 98 | 60 | Energy Environ. Sci., 2016, 9,2789 |
| **Pt-SACs MoSe$_2$ [9]** | 29 | 41 | Sci. Adv. 2018;4:eaao6657 |
| **Mo-SACs NC [10]** | 132 | 86 | Angew.Chem. Int. Ed. 2017, 56, 16086–16090 |
| **Pt SACs- NiO/Ni [11]** | 26 | 27.07 | Nat Commun 11, 1029 (2020) |
| **Co-SACs/PCN [12]** | 89 | 52 | Nat Commun 12, 3783 (2021) |
| **Pt-DG [13]** | 37 | 119 | Nat Catal 2, 134–141 (2019) |
| **Pt-SAs PCM [14]** | 50 | 73.6 | J.Am.Chem.Soc.2022,144, 2171−2178 |
| **Pt-SACs NC [15]** | 46 | 36.8 | Sci.Adv. 2018;4:eaao6657 |
| **Co-SACs β-Mo$_2$C** [16] | 141 | 62 | Adv. Funct. Mater., 2020, 30, 2000561. |
| **P/FeCo-NC** [17] | 95 | 72 | Adv. Energy Mater. 2024, 2404167 |
| **Ni$_5$P$_4$-Ru** [18] | 54 | 52 | Adv. Mater. 32, 1906972 (2020) |
| **Ni-SAs-PtNWs** [19] | 70 | 60.3 | Nat. Catal. 2, 495-503 (2019) |
| **Pt/np-Co0.85Se** [20] | 58 | 39 | Nat. commun. 10, 1743 (2019) |
| **Co-P$_2$N$_2$-C** [21] | 38 | 66.9 | Nat.Commun 15, 2774 (2024) |

# Reference

1 Feng, X. *et al.* Cu/CuxO/Graphdiyne Tandem Catalyst for Efficient Electrocatalytic Nitrate Reduction to Ammonia. *Advanced Materials* **36**, 2405660, doi:https://doi.org/10.1002/adma.202405660 (2024).

2 Zhou, K. *et al.* Seamlessly conductive Co(OH)2 tailored atomically dispersed Pt electrocatalyst with a hierarchical nanostructure for an efficient hydrogen evolution reaction. *Energy and Environmental Science* **13**, 3082-3092 (2020).

3 Yang, W. *et al.* Tuning the Cobalt–Platinum Alloy Regulating Single-Atom Platinum for Highly Efficient Hydrogen Evolution Reaction. *Advanced Functional Materials* **32**, 2205920, doi:https://doi.org/10.1002/adfm.202205920 (2022).

4 Yan, Y. *et al.* Atomic-Level Platinum Filling into Ni-Vacancies of Dual-Deficient NiO for Boosting Electrocatalytic Hydrogen Evolution. *Advanced Energy Materials* **12**, 2200434, doi:https://doi.org/10.1002/aenm.202200434 (2022).

5 Zhang, H., Zhou, W., Lu, X. F., Chen, T. & Lou, X. W. Implanting Isolated Ru Atoms into Edge-Rich Carbon Matrix for Efficient Electrocatalytic Hydrogen Evolution. *Advanced Energy Materials* **10**, 2000882, doi:https://doi.org/10.1002/aenm.202000882 (2020).

6 Wei, J. *et al.* In situ precise anchoring of Pt single atoms in spinel Mn3O4 for a highly efficient hydrogen evolution reaction. *Energy & Environmental Science* **15**, 4592-4600, doi:10.1039/D2EE02151J (2022).

7 Jiang, P. *et al.* Pt-like electrocatalytic behavior of Ru–MoO2 nanocomposites for the hydrogen evolution reaction. *Journal of Materials Chemistry A* **5**, 5475-5485, doi:10.1039/C6TA09994G (2017).

8 Zhang, J. *et al.* Engineering water dissociation sites in MoS2 nanosheets for accelerated electrocatalytic hydrogen production. *Energy & Environmental Science* **9**, 2789-2793, doi:10.1039/C6EE01786J (2016).

9 Shi, Y. *et al.* Electronic metal–support interaction modulates single-atom platinum catalysis for hydrogen evolution reaction. *Nature Communications* **12**, 3021, doi:10.1038/s41467-021-23306-6 (2021).

10 Chen, W. *et al.* Rational Design of Single Molybdenum Atoms Anchored on N-

Doped Carbon for Effective Hydrogen Evolution Reaction. *Angewandte Chemie International Edition* **56**, 16086-16090, doi:https://doi.org/10.1002/anie.201710599 (2017).

11  Zhou, K. L. *et al.* Platinum single-atom catalyst coupled with transition metal/metal oxide heterostructure for accelerating alkaline hydrogen evolution reaction. *Nature Communications* **12**, 3783, doi:10.1038/s41467-021-24079-8 (2021).

12  Cao, L. *et al.* Identification of single-atom active sites in carbon-based cobalt catalysts during electrocatalytic hydrogen evolution. *Nature Catalysis* **2**, 134-141, doi:10.1038/s41929-018-0203-5 (2019).

13  Yang, Q. *et al.* Single Carbon Vacancy Traps Atomic Platinum for Hydrogen Evolution Catalysis. *Journal of the American Chemical Society* **144**, 2171-2178, doi:10.1021/jacs.1c10814 (2022).

14  Zhang, H. *et al.* Dynamic traction of lattice-confined platinum atoms into mesoporous carbon matrix for hydrogen evolution reaction. *Science Advances* **4**, eaao6657, doi:doi:10.1126/sciadv.aao6657 (2018).

15  Fang, S. *et al.* Uncovering near-free platinum single-atom dynamics during electrochemical hydrogen evolution reaction. *Nature Communications* **11**, 1029, doi:10.1038/s41467-020-14848-2 (2020).

16  Ma, Y. *et al.* Synergistically Tuning Electronic Structure of Porous β-Mo2C Spheres by Co Doping and Mo-Vacancies Defect Engineering for Optimizing Hydrogen Evolution Reaction Activity. *Advanced Functional Materials* **30**, 2000561, doi:https://doi.org/10.1002/adfm.202000561 (2020).

17  Tao, X. *et al.* Phosphorus-Enhanced Bimetallic Single-Atom Catalysts for Hydrogen Evolution. *Advanced Energy Materials* **n/a**, 2404167, doi:https://doi.org/10.1002/aenm.202404167.

18  He, Q. *et al.* Achieving Efficient Alkaline Hydrogen Evolution Reaction over a Ni5P4 Catalyst Incorporating Single-Atomic Ru Sites. *Advanced Materials* **32**, 1906972, doi:https://doi.org/10.1002/adma.201906972 (2020).

19  Li, M. *et al.* Single-atom tailoring of platinum nanocatalysts for high-performance multifunctional electrocatalysis. *Nature Catalysis* **2**, 495-503, doi:10.1038/s41929-

019-0279-6 (2019).

20  Jiang, K. *et al.* Single platinum atoms embedded in nanoporous cobalt selenide as electrocatalyst for accelerating hydrogen evolution reaction. *Nature Communications* **10**, 1743, doi:10.1038/s41467-019-09765-y (2019).

21  Qian, S. *et al.* Tailoring coordination environments of single-atom electrocatalysts for hydrogen evolution by topological heteroatom transfer. *Nature Communications* **15**, 2774, doi:10.1038/s41467-024-47061-6 (2024).